\documentclass[prl,twocolumn,showpacs,preprintnumbers,aps,nofootinbib]{revtex4-1}
\usepackage{graphicx}
\usepackage{dcolumn}
\usepackage{bm}
\usepackage{amsmath,amssymb,amsfonts}
\usepackage{latexsym}
\usepackage{bbm}
\usepackage{color}

\begin{document}


\title{
 Approximate Alignment 
 in Two Higgs Doublet Model with Extra Yukawa Couplings
 }

\author{Wei-Shu Hou and Mariko Kikuchi}
\affiliation{Department of Physics, National Taiwan University, Taipei 10617, Taiwan}
\bigskip

\date{\today}

\begin{abstract}
With discovery of the 125 GeV boson $h^0$, the existence of
a second doublet is very plausible.
We show that the ``alignment'' phenomenon, that
$h^0$ is found to resemble closely the Standard Model Higgs boson,
may correspond to Higgs quartic couplings $\eta_i$
that are ${\cal O}(1)$ in strength.
If the exotic bosons of the second doublet
possess extra top Yukawa couplings, which are the least constrained by data,
such a two Higgs doublet model could drive electroweak baryogenesis,
as well as further ``protect'' the apparent alignment.
The exotic Higgs bosons can be sub-TeV in mass
while remaining well hidden so far, with broad parameter space
for search at the Large Hadron Collider.
%
\end{abstract}


\maketitle

\paragraph{Introduction.---}

The 125 GeV scalar boson $h^0$ discovered~\cite{h125_discovery} in 2012
resembles closely~\cite{Khachatryan:2016vau}
the Higgs boson of the Standard Model (SM),
but no sign of physics beyond SM has emerged
so far at the Large Hadron Collider (LHC). 
This includes supersymmetry (SUSY), where its scale
could be~\cite{Craig, Athron:2017yua} at several TeV,
with some light state(s) perhaps still elusive.

The $h^0$ boson belongs to the weak doublet $\Phi$ that
has a vacuum expectation value (VEV) $v \cong 246$~GeV,
inducing electroweak symmetry breaking (EWSB).
Since all weak fermion doublets come in three copies,
it is reasonable that a second scalar doublet $\Phi'$ exists,
and the physical $H^0$, $A^0$ and $H^\pm$ bosons
should be pursued at the LHC. But where are they?
In fact, SUSY implies two Higgs doublets,
%
but a recent minimal SUSY (MSSM) fit
places the exotic scalars at 5 TeV~\cite{Athron:2017yua},
which is in the decoupling limit~\cite{Gunion:2002zf},
and would be bad news for {the} LHC search.
It echoes also the apparent ``alignment'',
the observed proximity~\cite{Khachatryan:2016vau}
of $h^0$ to the SM Higgs boson.
The MSSM fit, however, is actually quite
``flat''~\cite{MWhite} in $m_{\Phi'}$, and in any case
does not stop phenomenological studies~\cite{Craig:2013hca, Carena:2013ooa,
 Carena:2014nza, Haber:2015pua, Bernon:2015qea, Bechtle:2016kui, Han:2017pfo}
from probing below the TeV scale.

%
Most phenomenological and experimental studies for the LHC
are in the SUSY-inspired two Higgs doublet model (2HDM),
called 2HDM-II, where up- and down-type quarks receive
mass from separate scalar doublets.
In this way, mass and Yukawa matrices are diagonalized simultaneously,
and flavor changing neutral Higgs (FCNH) couplings are absent.
This is the Natural Flavor Conservation (NFC) criteria~\cite{Glashow:1976nt}
of Glashow and Weinberg, where a $Z_2$ symmetry is invoked
for the Higgs and fermion fields.
%
As such, 2HDM-II (and also the so-called 2HDM-I)
has no new Yukawa couplings.

It is, however, the unique $CP$ violating (CPV) phase
in quark mixing, rooted in Yukawa couplings,
that accounts for all laboratory-observed CPV phenomena.
Given the deep shortfall from explaining the
baryon asymmetry of the Universe (BAU),
it is natural to ask whether extra Yukawa couplings
could be considered without the usual $Z_2$.
It was noted long ago~\cite{Cheng:1987rs} that,
given the trickle-down pattern or mass suppression of
off-diagonal quark mixings, NFC
(hence the $Z_2$ symmetry) may not be necessary,
while $t\to ch$ decay induced by the least constrained
$tch$ coupling could serve as hallmark~\cite{Hou:1991un}.
After {the} $h^0$ discovery, it was further advocated~\cite{Chen:2013qta}
that, much like flavor couplings within SM,
one should just let Nature decide on the FCNH couplings.

Starting from {the} general 2HDM without the $Z_2$ symmetry,
we show that approximate alignment may reflect
${\cal O}(1)$ couplings in the Higgs potential.
Such couplings could bring about first order
electroweak phase transition, which
in fact could work also for 2HDM-II. 
But the bonus~\cite{Fuyuto:2017ewj} for the general 2HDM
is that it could attain electroweak baryogenesis (EWBG),
i.e. generate BAU, where
\emph{new} ${\cal O}(1)$ top quark Yukawa couplings
provide new sources of CPV.
We argue that NFC protection against FCNH can be replaced
by approximate alignment, together with a flavor organizing
principle reflected in SM itself.
Interestingly, the ${\cal O}(1)$ extra top Yukawa coupling
may help protect alignment~\cite{Hou:2017vvp}.

\paragraph{2HDM-II: Alignment 
 without Decoupling.---}

In want of additional scalar bosons at sub-TeV scale,
the issue of ``alignment without decoupling'' has certainly
been a topic of discussion~\cite{Gunion:2002zf, Craig:2013hca, Carena:2013ooa,
 Carena:2014nza, Haber:2015pua, Bernon:2015qea, Bechtle:2016kui, Han:2017pfo}.
Let us define alignment more precisely: in notation of 2HDM-II,
near alignment means the mixing angle $\cos(\beta - \alpha)$
between $h^0$ and the heavy $CP$-even scalar $H^0$ is small,
where $\tan\beta = v_2/v_1$ is the ratio of VEVs of the two doublets,
distinguished by the $Z_2$ symmetry that enforces NFC,
while $\alpha$ is a mixing parameter arising from the Higgs potential.

It is useful to clarify some confusion brought about
by this notation. Given that $\beta$ and $\alpha$ have
rather different origins, one may ask:
Why should $\beta \simeq \alpha \pm\pi/2$?
%
This is exacerbated by claimed ``solutions" for
alignment without decoupling.
%
%
For example, whether {in the form of Eq.~(4.8) or Eq.~(4.14)}
of Ref.~\cite{Carena:2013ooa}, the {proposed} solutions
equate $\tan\beta$ with some combination of quartic Higgs couplings,
be it from Higgs potential, or from loop effects.
One has alignment without decoupling
if the equality holds approximately, but
the question {then}
is why would $\tan\beta$ match the right-hand side.
Efforts have been made to elucidate~\cite{Carena:2013ooa, Carena:2014nza, Bechtle:2016kui}
the situation in MSSM, where one needs fine-tuned 
cancellations between tree level and up to two-loop
corrections~\cite{Bechtle:2016kui} from top (squark) sector.
As we will discuss shortly, $\beta - \alpha$ should be
viewed as a single angle between the so-called
Higgs basis~\cite{Botella:1994cs} and
neutral Higgs mass basis~\cite{HHaber}.

The sense of ``fine-tuning'' might reflect some underlying symmetry.
Ref.~\cite{Dev:2014yca} found three maximal symmetries
(with SO(5) the simplest) of the Higgs potential for two Higgs doublets,
that ``naturally realize the alignment limit'' without need of large $\tan\beta$,
hence may be an improvement.
Renormalization group effects would break this symmetry,
hence deviate from alignment.
In this Letter, we take a simpler 
approach:
rather than more symmetry that may link to high scales,
we focus on the Higgs potential and remove all symmetry,
%
which is opposite the direction of exploration from Ref.~\cite{Dev:2014yca}.
%
%
%
In the following, we explore the Higgs sector,
and return to flavor and phenomenology issues subsequently.

\paragraph{General Higgs Potential and Alignment.---}

The Higgs potential of the 2HDM without 
$Z_2$ symmetry is,
\begin{align}
 &V(\Phi,\, \Phi')
= \mu_{11}^2 |\Phi|^2 +\mu_{22}^2 |\Phi'|^2
        - \left(\mu_{12}^2 \Phi^\dagger \Phi' + {\rm h.c.}\right) \notag\\
   &\ +\frac{\eta_1}{2}|\Phi|^4 + \frac{\eta_2}{2}|\Phi'|^4 + \eta_3|\Phi|^2|\Phi'|^2
   + \eta_4|\Phi^\dagger \Phi'|^2 \notag\\
   &\ +\left\{\frac{\eta_5^{}}{2}(\Phi^\dagger \Phi')^2
   + \left[\eta_6^{}|\Phi|^2 + \eta_7^{}
      |\Phi'|^2\right]\Phi^\dagger \Phi' + {\rm h.c.}\right\}.
 \label{eq:pote2}
\end{align}
The usual notation for Higgs potential with $Z_2$ symmetry,
where $\Phi \to \Phi$ and $\Phi' \to -\Phi'$ under $Z_2$
is in form of $m^2_{ij}$ and $\lambda_i$, with $\lambda_6 = \lambda_7 = 0$.
We take $V$ as $CP$ conserving,
hence all parameters are real for simplicity.
Without $Z_2$, $\Phi$ and $\Phi'$ cannot be distinguished
(and $\tan\beta$ is unphysical), we can \emph{choose}
the (Higgs) basis where $\Phi$ generates $v$ for EWSB,
with minimization conditions~\cite{Davidson:2005cw, Haber:2006ue, Haber:2010bw}
\begin{align}
  \mu_{11}^2 = -\frac{1}{2}\eta_1^{}v^2, \quad
  \mu_{12}^2 =  \frac{1}{2}\eta_6^{}v^2.
 \label{eq:minim}
\end{align}
%
{The second condition reduces
the parameter count by one.}
%
To be explicit,
Eq.~(1) has the parameters
 $v$, $\mu_{22}^2$ (positive definite) and $\eta_i$ with $i = 1$--7;
with $Z_2$, the parameters are
 $v$, $\tan\beta$, $m_{12}^2$ (soft breaking) and $\lambda_i$ with $i = 1$--5.
As the Higgs basis \emph{always} exist, a relation
 (Eq.~(54) of Ref.~\cite{Haber:2015pua})
between $\eta_i$'s must be satisfied to guarantee
one could rotate to the ``$Z_2$ basis''~\cite{Haber:2015pua}
where $\lambda_6 = \lambda_7 = 0$,
through which the $\tan\beta$ value can be found~\cite{HHaber}.
Thus, our following discussion
{\emph{applies also to 2HDM-II}}.

The charged and $CP$-odd Higgs masses are,
\begin{align}
  m_{H^\pm}^2 &= \mu_{22}^2 + \frac{1}{2}\eta_3^{}v^2, \\
  m_A^2 & = \mu_{22}^2 +\frac{1}{2}(\eta_3 +\eta_4^{} - \eta_{5}^{})v^2.
\end{align}
For the $CP$-even Higgs mass matrix, one has
\begin{align}
  M_\textrm{even}^2 =
  \left[\begin{array}{cc}
    \eta_1^{}v^2 & \eta_6^{} v^2 \\
    \eta_6^{} v^2 & \mu_{22}^2 + \frac{1}{2} (\eta_3^{} + \eta_4^{} + \eta_5)v^2\\
    \end{array}\right],
\end{align}
which is diagonalized by
\begin{align}
  R^T_\gamma M_\textrm{even}^2 R_\gamma  =
    \left[\begin{array}{cc}
    m_H^2 & 0 \\
    0 & m_h^2 \\
  \end{array}\right], \quad
  R_\gamma  =  \left[\begin{array}{cc}
    c_\gamma & - s_\gamma \\
    s_\gamma & c_\gamma \\
  \end{array}\right],
\end{align}
with convention similar to 2HDM-II:
$c_\gamma \equiv \cos\gamma$ replaces $\cos(\beta -\alpha)$,
and $s_\gamma \equiv \sin\gamma$.
%
It is now clear that $\gamma \equiv \beta - \alpha$
is the relative angle between the Higgs basis
and the neutral Higgs mass basis,
and is basis-independent.

Taking $m_h \cong 125$ GeV the observed value, we do
not give the detailed formula for $m_H$,
but note the mixing angle $c_\gamma$ satisfies two relations,
\begin{align}
   c_\gamma^2  = \frac{\eta_1^{}{v^2} - m_h^2}{m_H^2 - m_h^2}, \quad 
  \sin2\gamma^{}  = \frac{2\eta_6^{}v^2}{m_H^2 - m_h^2}. \label{eq:npara}
\end{align}
In alignment \emph{limit} of $c_\gamma \to 0$, $s_\gamma \to -1$,
one has $\eta_1 \to m_h^2/v^2 \simeq 0.26$.
For $c_\gamma$ small but nonvanishing, $\eta_1 v^2 - m_h^2$ 
can be weighed down by $m_H^2 - m_h^2$. 
Since $s_\gamma \to -1$ always holds better
than $c_\gamma \to 0$, the second relation gives
\begin{align}
  c_\gamma^{} & \simeq \frac{-\eta_6^{}v^2}{m_H^2 - m_h^2},
  \ \ \ {\rm (near\; alignment)} \label{eq:Haber}
\end{align}
which appears e.g. in
{Refs.~\cite{Bernon:2015qea, Bechtle:2016kui}.} 

{The discussion in Ref.~\cite{Bernon:2015qea}
is more general (see below), but both
Refs.~\cite{Bernon:2015qea, Bechtle:2016kui}
focused on $|\eta_6| \ll 1$ in their discussions.
We stress, however, }
that small $\eta_6$ is in general not needed
in {the} 2HDM: 
$c_\gamma$ can be small for
\begin{align}
  |\eta_6| \sim {\cal O}(1),\ \ m_H^2 - m_h^2 \gtrsim {\rm several}\; v^2.  \label{eq:alignment}
\end{align}
Note that a low $m_h^2/v^2 \simeq 0.26$
is \emph{not} required,
i.e. $c_\gamma$ can be small even for $m_h \sim 300$ GeV,
so long that $m_H^2 - m_h^2 \gtrsim {\rm several}\; v^2$.
%
%
%
{But $m_H^2 - m_h^2 \gtrsim {\rm several}\; v^2$
implies that at least one of $\eta_3$, $\eta_4$, $\eta_5$ is ${\cal O}(1)$,
or (inclusive) $\mu_{22}^2 > v^2$.}
Furthermore, for $\eta_6 \sim {\cal O}(1)$,
level repulsion could drive $m_H^2 - m_h^2$ apart and
push $m_h^2/v^2$ down from $\eta_1 \sim {\cal O}(1)$.
On the other hand, given the observed $m_h^2/v^2 \sim 0.26$,
if one allows the tuning of $\eta_6 \lesssim 1/4$,
one could get \emph{extreme} alignment, i.e. $c_\gamma \to 0$,
resulting in $\eta_1 \to 0.26$.

But in general, $\eta_6$ need not be small!
This can in fact be seen from Fig.~1 of Ref.~\cite{Bernon:2015qea},
the numerical study of 2HDM-II and 2HDM-I:
$|Z_6| \sim 1$ can still give $|c_{\beta - \alpha}|
\sim 10\%$ (or more) for $m_H \lesssim$ TeV,
where $Z_6$ is equivalent to our $\eta_6$.
However, {as mentioned},
this study focused on very small $|Z_6|$ values,
and $|Z_6| \sim 1$ was barely discussed.

\begin{figure*}[t]
\center
\includegraphics[width=6.1cm]{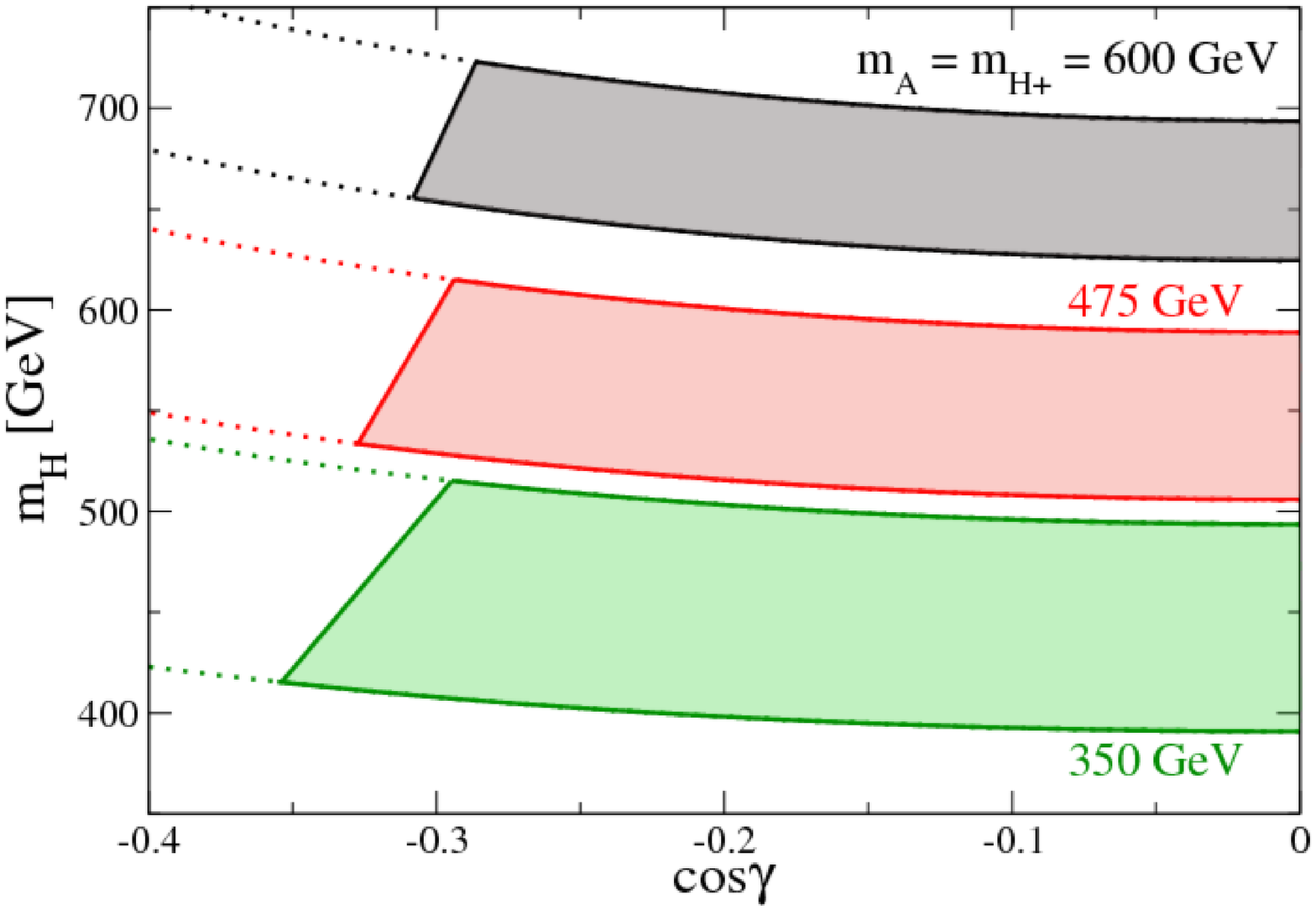} \hskip0.3cm
\includegraphics[width=6cm]{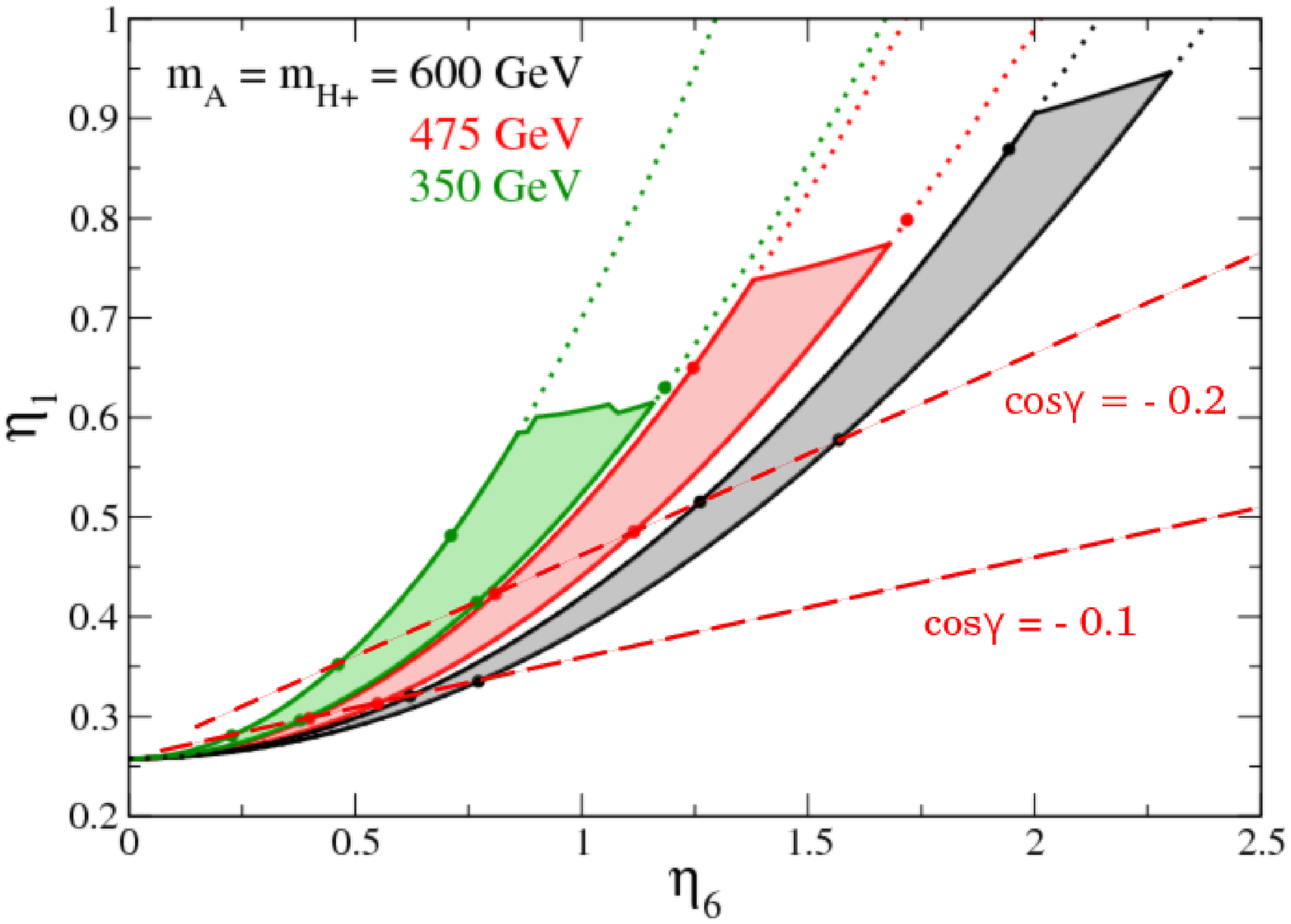}
\caption{
For $m_A = m_{H^+} = 350$, 475, 600 GeV
and $\eta_4 = \eta_5 \equiv \eta' \in (0.5,\; 2)$:
[left] $m_H$ vs $\cos\gamma$;
[right] $\eta_1$ vs $\eta_6$, where filled circles are
 for $-\cos\gamma = 0.1,\ 0.2,\ 0.3$.
The 95\% C.L. limit from $S$-$T$ data~\cite{Baak:2012kk}  cuts off
parameter space (dotted lines).
}
\label{fig:fig1}
\end{figure*}

\paragraph{Numerical Illustration with Custodial SU(2).---}

Our presentation so far has been general,
and as we have just commented, does not really
depend on whether a $Z_2$ symmetry is imposed or not.
But with $\eta_{i} \sim {\cal O}(1)$, 
one should check precision electroweak constraints.
There are two corrections to the $T$ parameter,
\begin{align}
  \Delta T^{\rm 2HDM} = \Delta T_{SS} + \Delta T_{SV},
  \label{eq:DeltaT}
\end{align}
where the $SS$ term involves only scalars, 
while the $SV$ term involves both scalars and vectors.
We note that
\begin{align}
  \eta_4 = \eta_5 \equiv \eta', \quad ({\rm custodial\ SU(2)})
  \label{eq:custodial}
\end{align}
gives $m_A = m_{H^+}$ hence restores
the custodial SU(2) symmetry~\cite{Pomarol:1993mu, custodial},
which can be verified through $V(\Phi,\, \Phi')$.
Assuming 
Eq.~(\ref{eq:custodial}), the 
$\Delta T_{SS}$ term vanishes.
However, $\Delta T_{SV}$ does not vanish because $M_W \neq M_Z$,
and provides some constraint as it is proportional to $c_\gamma^2$.

Taking the $T$ parameter constraint into account,
we illustrate approximate alignment for the simplified case of
custodial SU(2), i.e. $m_A^2 = m_{H^+}^2$.
We plot $m_H$ vs $c_\gamma^{}$ in Fig.~1[left]
for fixed $m_A = m_{H^+} = 350$, 475, 600 GeV,
and for $\eta'$ (or $\eta_4 = \eta_5$) ranging from 0.5 to 2.
We see that the 95\% C.L. limit from $S$-$T$ parameter data~\cite{Baak:2012kk}
cuts the solution space off around $-c_\gamma^{} \sim 0.3$,
illustrated by solid lines that turn into dotted lines.
This corresponds to $-s_\gamma^{} \simeq 0.95$,
which is still close to alignment.
To be conservative, one can take $-c_\gamma < 0.2$,
or $-s_\gamma > 0.98$, which is rather close to {the} alignment limit.

In Fig.~1[right] we plot $\eta_1$ vs $\eta_6$, where
$-c_\gamma^{} = 0.1,\ 0.2,\ 0.3$
 ($-s_\gamma \cong 0.995,\ 0.980,\ 0.954$)
are marked by filled circles.
We see that, for $m_A = m_{H^+} =350$, 475, 600 GeV,
one has $-c_\gamma \lesssim 0.2$ for $\eta_6 \lesssim 0.5$, 1, 1.5,
respectively, i.e. close to alignment,
while for $\eta_6 \lesssim m_h^2/v^2 \cong 0.26$,
one is \emph{very} close to {the} alignment limit.
%
%
We note further that, while $\eta_1 < \eta_6$ in general,
there is large parameter space where both are ${\cal O}(1)$,
and \emph{$m_h^2$ is pushed down from larger $\eta_1 v^2$ values
to what is observed by level repulsion}.
We stress that Fig.~1 is an illustration, not yet a scan.

Although $m_A = m_{H^+} = 350$, 475, 600 GeV
do not distinguish between $\mu_{22}^2$ and $\eta_3 v^2/2$,
they imply
\begin{align}
  \mu_{22}^2/v^2 + \eta_3/2 \simeq 2,\ 3.7,\ 6,
  \label{eq:param}
\end{align}
respectively. If one assumes equal share between
the two terms, one has $\eta_3 \sim 2$ for 350 GeV,
while for 600 GeV, $\eta_3 \sim 6$ is becoming sizable.
If we keep $\eta_3 \lesssim 3$ for perturbativity,
then $\mu_{22}^2 \gtrsim (367\ {\rm GeV})^2,\  (519\ {\rm GeV})^2$,
respectively, for $m_A = m_{H^+} = 475$, 600 GeV.
Thus, although our alignment discussion is not affected,
the inertial 
$\mu_{22}^2$
would take up more role for $m_A$, $m_{H^+} \gtrsim 500$ GeV.
{In this context we remark that,
just having one $|\eta_i| \sim 1$ implies
the Landau pole is not far above 10 TeV.
Thus, ${\cal O}(1)$ Higgs quartic couplings imply
a nearby strong coupling scale, which would be
a boon to a 100 TeV $pp$ collider.
Some Higgs quartic couplings could be negative, however,
hence the Landau pole scale is rather uncertain.}

\begin{figure*}[t]
\center
\includegraphics[width=17cm]{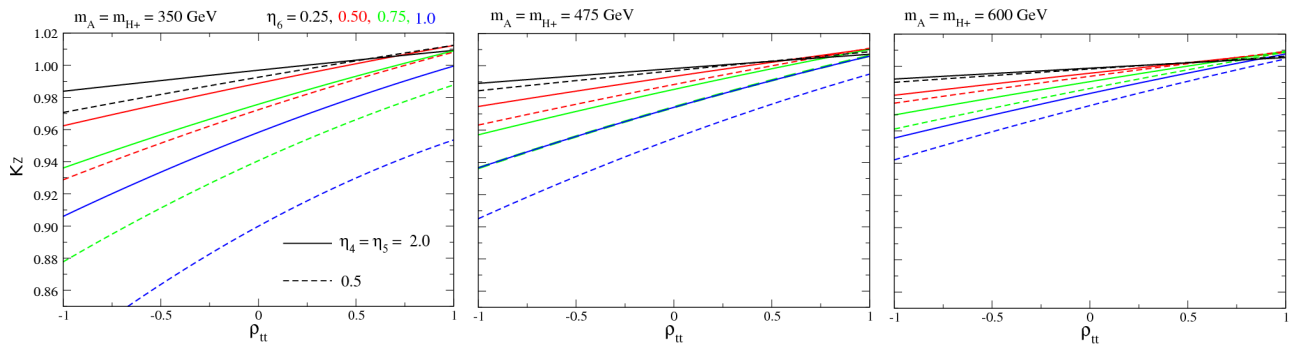}
\caption{
Same as Fig.~1, but for $\kappa_Z$ ($\sim$
$|\Gamma_{h \to ZZ^*}/\Gamma_{h \to ZZ^*}^{\rm SM}|^{1/2}$ measured experimentally)
vs $\rho_{tt}$ for several $\eta_6 \leq 1$ values.
}
\label{fig:fig2}
\end{figure*}

\paragraph{One-loop Protection.---}
%

%
So far our discussion does not depend on imposing
a $Z_2$ symmetry or not.
We wish, however, to further point out
a mechanism~\cite{Hou:2017vvp}, within 2HDM without $Z_2$,
where alignment could be ``protected'' at loop level.
With all $\eta_i \sim {\cal O}(1)$, the one loop
corrections to, e.g. $\Gamma(h \to ZZ^*)$ by the extra scalars
are significant, resulting in reduction.
But while Yukawa couplings of $\Phi$ (Eq.~(\ref{eq:pote2})),
the mass-giving doublet, are diagonalized with fermion masses,
those of $\Phi'$ are not,
and the point~\cite{Chen:2013qta} of general 2HDM is that
Nature has some \emph{flavor organizing principle} that
suppresses FCNH involving low mass fermions.
Given $\lambda_t \cong 1$, however, the top-related extra Yukawa couplings
$\rho_{tt}$ and $\rho_{ct}$ could be ${\cal O}(1)$~\cite{Chen:2013qta}.
We have shown~\cite{Hou:2017vvp} that, for $\rho_{tt} c_\gamma < 0$,
the top loop effect enhances $\Gamma(h \to ZZ^*)$,
and could compensate a negative deviation due to bosonic loops.
Thus, approximate alignment may be only apparent,
although 
$\rho_{tt} c_\gamma$ of wrong sign
would be in {the} wrong direction.

In Fig.~2 we plot $\kappa_Z$, 
the one-loop corrected $hZZ$ coupling~\cite{Hou:2017vvp}
in the general 2HDM normalized to SM,
vs $\rho_{tt}$ for $m_A = m_{H^+}$ as before,
and for $\eta' = 0.5$, 2 and several $\eta_6 \leq 1$ values.
As $c_\gamma < 0$, we see that $\rho_{tt} > 0$
could bring $\kappa_Z$ back to 1, but for $\rho_{tt} < 0$,
the top and bosonic loops together can push
$\kappa_Z$ sufficiently below SM expectation,
which is not observed~\cite{Khachatryan:2016vau}.
Note that for heavier $H^0$, the $\rho_{tt} < 0$
case is still allowed at present.
We have checked that the trend for $h \to \gamma\gamma$ is similar,
but because the experimental errors~\cite{Khachatryan:2016vau}
are more forgiving, they do not constrain the plots in Fig.~2.

Thus, 2HDM without $Z_2$ can not only give rise to small $|c_\gamma|$,
but provide a loop protection mechanism.

\paragraph{Phenomenology.---}

For the phenomenology of 2HDM without $Z_2$ symmetry,
one first needs to address
the recent bound~\cite{Misiak:2017bgg} of $m_{H^+} \gtrsim 580$ GeV.
This holds actually for 2HDM-II, as 
the $H^+$ effect is always~\cite{bsga2HDM} constructive with SM
in the $b\to s\gamma$ decay amplitude,
with leading piece independent of $\tan\beta$.
Thus, with the Belle update~\cite{Belle:2016ufb} of $B \to X_s\gamma$ being
slightly below SM expectation, a more stringent limit follows.
In the 2HDM without $Z_2$, however, there is no $\tan\beta$,
while the new Yukawa couplings $\rho_{ct}$, $\rho_{tt}$ (even $\rho_{cc}$)
and $\rho_{bb}$ enter the loop~\cite{Chen:2013qta},
with $\rho_{bb}$ constrained to be small
($|\rho_{bb}| \lesssim \lambda_b$, which is reasonable)
because of chiral enhancement.
These couplings are further complex, 
hence 
$b \to s\gamma$ carves out a solution space in the aforementioned parameters,
but the 580 GeV bound on $m_{H^+}$ valid for 2HDM-II
does not apply.

With $\tan\beta$ removed as a parameter,
the extra Higgs boson search strategy should be revised.
For example, $H^0/A^0 \to \tau^+\tau^-$ and $b\bar b$ search
are in the wrong direction, as the extra Yukawa couplings
should be rather small.
With $\rho_{tt} \sim {\cal O}(1)$, however,
the neutral states $A^0$ and $H^0$ are produced via gluon fusion
and decay strongly to $t\bar t$,
which is hard to disentangle~\cite{Frederix:2007gi, Carena:2016npr}
from the large $gg \to t\bar t$ background, due to
distortion from interference and detector smearing.
The ATLAS study~\cite{ATLAS:2016pyq} with 8 TeV data
takes interference effects into account,
but leverage is mainly for $\tan\beta \lesssim 0.7$ in 2HDM-II,
i.e. with enhanced top couplings, so
our $\rho_{tt} \sim {\cal O}(1)$ parameter space is not yet constrained.
Likewise, the ATLAS study of $pp \to tbH^{\pm}$ ($H^+ \to t\bar b$)~\cite{ATLAS:2016qiq}
at 13 TeV, or $t\bar t$ final states with additional heavy flavor jets~\cite{ATLAS:2016btu},
are again sensitive to low $\tan\beta$ values and do not yet constrain our scenario.
But with fast accumulation of Run 2 data, further discernment by ATLAS and CMS
are encouraged.

To account for {the} absence of signal for sub-TeV scalar bosons,
we have therefore chosen our lowest mass, $m_A = m_{H^+} = 350$ GeV,
to be above the $t\bar t$ threshold, which should be revisited by experiment.
Associated production, such as $t\bar t H^0/A^0$
and $\bar tbH^+(t\bar bH^-)$ followed by $H^0/A^0 \to t\bar t$
and $H^+ \to t\bar b(H^- \to \bar tb)$ decays
would be the search strategy~\cite{ATLAS:2016qiq, ATLAS:2016btu, Craig:2015jba}
at high luminosity.
%
An intriguing alternative, as studied in Ref.~\cite{Altunkaynak:2015twa},
may be $gg \to H^0/A^0 \to t\bar c(\bar tc)$ with decay mediated by
$\rho_{ct}$ coupling.
Despite being suppressed by $c_\gamma^2$,
search for $t \to ch^0$~\cite{Chen:2013qta} should certainly continue.
Note that $A^0 \to h^0 Z^0$, $H^\pm \to h^0W^\pm$ and $H^0 \to h^0h^0$ decays
are suppressed by $c_\gamma^2$, and can be ignored above
$t\bar t$ threshold.

Close to alignment and with custodial SU(2),
$m_H^2 > m_A^2 = m_{H^+}^2$ by $\eta' v^2$
{(see Eq.~(11)).}
So for $\eta' \lesssim (2m_Zm_A + m_Z^2)/v^2$,
one has $m_H - m_A \lesssim m_Z$.
Thus, for $m_A = 350$, 475 GeV,
$m_H - m_{A,\, H^+} \gtrsim m_Z$ for $\eta' \gtrsim 1.19$, 1.57,
while $m_H - m_A < m_Z$ for $m_A = 600$ GeV, unless $\eta' \gtrsim 2$.
Thus, for lower $m_A = m_{H^+}$ that enjoy larger cross sections,
there are interesting $H^0 \to A^0Z$, $H^\pm W^\mp$ decays.
The former could lead to $t\bar t Z$ signature,
with potential mass peaks in both $m_{t\bar t}$ and $m_{t\bar tZ}$.
The $H^\pm W^\mp \to t\bar bW^-(\bar tbW^+)$ final state,
however, may be swamped by $t\bar t$ background.
In any case, these processes would likely be subdominant,
unless $\rho_{tt}$ turns out to be far less than 1.

After completion of the original work of this Letter,
the process $cg \to tS^0 \to tt\bar c$, $tt\bar t$
($S^0 = H^0,\, A^0$) was explored~\cite{Kohda:2017fkn}, which
proceed via the extra $\rho_{ct}$ and $\rho_{tt}$ couplings.
It was found that the same-sign top~\cite{Hou:1997pm}
or triple top signatures
can be probed already with full LHC Run~2 data,
in contrast with the need of very high luminosity for the four-top
signature~\cite{Dev:2014yca, ATLAS:2016qiq, ATLAS:2016btu, Craig:2015jba}.

The general 2HDM with its FCNH couplings has much impact on flavor physics
and CPV (FPCP), and our parameter range in fact
overlaps well with the possibility~\cite{Fuyuto:2017ewj} of EWBG.
We refer to Ref.~\cite{Fuyuto:2017ewj} for discussion of related FPCP phenomenology,
such as electron (and neutron) electric dipole moment,
$h \to \mu\tau$ and $\tau \to \mu\gamma$,
as well as corrections to $h \to \gamma\gamma$ and $\lambda_{hhh}$.

\paragraph{Discussion and Conclusion.---}

The parameter space for alignment in 2HDM 
is larger than we discussed.
Custodial SU(2), or $m_A = m_{H^+}$,
used to control $\Delta T$,
need not be required:
the $\Delta T$ constraint would just carve out
some ``slices'' in $m_{H}$, $m_A$ and $m_{H^+}$ space.
%
%
The {derived condition that some} $\eta_i \sim {\cal O}(1)$
echoes the need for first order EWPT,
while ${\rm Im}\, \rho_{tt}$ drives
{the} CPV source for EWBG~\cite{Fuyuto:2017ewj}.
But 
${\cal O}(1)$ couplings are also not strictly necessary.
%
%
A full account of the phenomenology 
associated with alignment is beyond the scope of this work.

We have argued from Eq.~(\ref{eq:param}) that,
to keep $\eta_3$ perturbative,
$\mu_{22}^2/v^2$ has to grow in value
for $m_H \gtrsim 500$ GeV,
which 
would eventually lead to decoupling.
Thus, for sake of EWBG, Fig.~1 reflects the
mass range of interest for exotic Higgs
search at {the} LHC.

As an example of different parameter values,
suppose we let $\eta' < 0$ in Eq.~(\ref{eq:custodial}).
For $m_A = m_{H^+} = 550$ GeV and $\eta' \in (-1,\, -1/4)$,
$\eta_6 \simeq 1/4$, we have $m_H \in (493,\, 537)$ GeV
and $c_\gamma \simeq -0.067$ to $-0.056$, still close to alignment.
More negative $\eta'$ values would move $m_H$ lower.
Another example is ``twisted custodial'' symmetry~\cite{Gerard:2007kn},
$m_{H^0}^2 = m_{H^+}^2$, where $\eta_4 + \eta_5 = 0$
instead of Eq.~(\ref{eq:custodial}).
The situation is not so different from Fig.~1:
$m_{A^0} > m_{H^0} = m_{H^+}$ for $\eta_4 > 0$,
and $m_{A^0} < m_{H^0} = m_{H^+}$ for $\eta_4 <0$,
which is more constrained by $\Delta T$ but less aligned.

Small Cabibbo mixing, $\sin\theta_C \lesssim 1/4$, 
as well as the full quark mass-mixing pattern,
are not understood.
In contrast, approximate alignment may
reflect ${\cal O}(1)$ Higgs quartic couplings.
In retrospect, NFC~\cite{Glashow:1976nt} was an overkill against FCNH.
Even the Cheng-Sher ansatz~\cite{Cheng:1987rs}
was only an intermediate step.
The exotic $H^0/A^0$ can have FCNH couplings, but
approximate alignment suppresses such couplings for $h^0$.
The observed mass-mixing pattern,
rooted in {the} Yukawa couplings of the SM-like doublet,
further modulates the strength of Yukawa couplings
(including FCNH) of the exotic doublet.
There is much to look forward to if such Higgs bosons are discovered.

In conclusion, approximate alignment arises
in the general two Higgs doublet model
if {some of the} dimensionless parameters
in the Higgs potential, {including
$\mu_{22}^2/v^2$, are ${\cal O}(1)$}.
There is no need to impose small parameters,
and it does not depend on whether
a $Z_2$ discrete symmetry is imposed or not.
%
%
This revitalizes the outlook for sub-TeV $H^0$, $A^0$ and $H^\pm$ bosons.
Discovery of such bosons with ${\cal O}(1)$ Higgs couplings
would point to a new scale below {10--20} TeV.
We have advocated the two Higgs doublet model without $Z_2$ symmetry,
as it offers hope for electroweak baryogenesis,
precisely for Higgs quartic and extra top Yukawa couplings
that are ${\cal O}(1)$ in strength.

%


\vskip0.2cm
\noindent{\bf Acknowledgments} \
We thank J.-M.~G\'erard, M.~Kohda, E.~Senaha, M. White
and T.~Yanagida, and especially H.E. Haber, for discussions.
This research is supported by grants MOST 104-2112-M-002-017-MY2,
105-2112-M-002-018, 106-2811-M-002-010,
and NTU 106R8811, 106R104022.
WSH is grateful to E.~Gardi and F.~Muheim
for a fruitful stay at University of Edinburgh.


\end{document}